\documentclass[prepring,showpacs,preprintnumbers,floatfix]{revtex4}
\usepackage{graphicx}
\usepackage{dcolumn}
\usepackage{bm}
\usepackage{longtable}

\begin{document}

\title{Structure of mesoscopic Coulomb balls}

\author{P.~Ludwig$^{1,2}$, S.~Kosse$^{1}$, and M.~Bonitz$^{1}$}
\affiliation{$^{1}$Christian-Albrechts-Universit{\"a}t zu Kiel,
Institut f{\"u}r Theoretische Physik und Astrophysik, Leibnizstr. 15, 24098 Kiel, Germany}
\affiliation{$^{2}$Universit{\"a}t Rostock, Fachbereich Physik, l8051 Rostock, Germany}
\date{\today}

\begin{abstract}
An analysis of the structural properties of three-dimensional Coulomb clusters
confined in a spherical parabolic trap is presented. Based on extensive high accuracy
computer simulations the shell configurations and energies  for particle numbers in
the range $60 \le N\le 160$ are reported. Further, the intrashell symmetry and the lowest
metastable configurations are analyzed for small clusters and a novel type of excited state
which does not involve a change of shell configuration is identified.
\end{abstract}
\maketitle

Spatially confined mesoscopic charged particle systems have a number of unique
properties not observed in conventional {\em quasineutral macroscopic} plasmas of electrons and
ions in discharges or solids, electrons and holes in highly excited semiconductors and so on.
With the help of confinement potentials it has now become routine to trap, for long
periods of time, {\em plasmas of a single charge} (nonneutral plasmas), e.g. electrons and ions
and even positrons in Paul and Penning traps \cite{walter87,wineland87}, for an overview see
\cite{dubin99}, or colloidal (dusty) plasmas in discharge chambers, e.g. \cite{morfill03}.
By varying the confinement strength researchers have achieved liquid behavior and even Coulomb
crystallization of ions \cite{itano98} and dust particles \cite{thomas94,hayashi94}.
These strong correlation phenomena are of exceptional current interest in a large variety of fields
ranging from astrophysics (interior of Giant planets), high-power laser compressed laboratory plasmas,
to condensed matter and quantum dots \cite{afilinov-etal.01prl} etc. Coulomb (Wigner)
crystals are expected to exist in many White Dwarf stars.

A particular property of trapped mesoscopic ($N\lesssim 1,000$) clusters in spherical traps is the
occurence of concentric shells with characteristic occupation numbers, shell closures and unusual stable
``magic'' configurations. Due to their close similarity to nuclei, metal clusters or atoms, these
mesoscopic systems have been called ``artificial atoms''. A number of papers
has been devoted to the exploration of the energetically lowest shell configuration (ground state) and
metastable (``excited'') states of two-dimensional (2D) artificial atoms, e.g.
\cite{bedanov94,kong03,ludwig03} and references therein.

Recently, Arp et al. succeeded in the first experimental creation of {\em spherically symmetric 3D clusters}, so-called ``dust Coulomb balls'' \cite{piel2004} and presented an almost perfect crystal
of $N=190$ dust particles.
This raises the question about theoretical configurations of mesoscopic 3D Coulomb balls
which is the subject of this paper. It is natural to start with an analysis of the ground state and
lowest metatstable states, referring finite temperature and melting properties, e.g. \cite{schiffer02},
to a subsequent study.

The theoretical analysis of 3D artificial atoms is much more involved than in 2D and has so far mostly
been restricted to small cluster sizes with often conflicting results, e.g. \cite{rafac91,hasse91,tsuruta93} and references therein.
Rafac et al. \cite{rafac91}, correcting earlier results, identified the first shell closure at $N=12$
(the 13th particle is the first to occupy a second shell) and presented detailed data, including ground state energies for $N \le 27$, but they missed the onset of the third shell, as did Hasse et al. \cite{hasse91}. Tsuruta et al. extended the table to $N=59$ \cite{tsuruta93}. The most extensive data, for up to a
few thousand particles, has been presented by Hasse et al. \cite{hasse91} and is a valuable reference for
theoretical and experimental groups. However, their tables report excited states rather than the ground states for $N=28-31, 44, 54$ and practically for all $N>63$ (except for $N=66$). The reason for the computational difficulties is the existence of a large number of excited (metastable) states
which are energetically close to the ground state; with increasing $N$ this number grows exponentially whereas the energy difference rapidly vanishes.
This has to be accounted for by the computational strategy and choice of
accuracy, see below.

{\em Model:}
We consider $N$ classicle particles with equal charge $q$ and mass $m$ interacting via the
Coulomb force and being confined in a 3D isotropic harmonic trap with frequency $\omega$
with the hamiltonian
\begin{eqnarray}\label{hamilton}
H_N =  \sum\limits_{i=1}^{N} \frac{m}{2}\dot{r_i}^2 +
         \sum\limits_{i=1}^{N} \frac{m}{2}\omega^2 r_i^2 +
	 \sum\limits_{i>j}^{N} \frac{q^2}{4\pi\varepsilon|{\bf r}_i-{\bf r}_j|}.
\end{eqnarray}
Despite its simplicity, model (\ref{hamilton}) captures the basic properties of a multitude of classical systems,
in particular of dust Coulomb balls and ions in spherical traps.
Below we will use dimensionless lengths and energies by introducing the units $r_0 = (q^2/2 \pi \varepsilon m \omega^2)^{1/3}$
and $E_0 = (m \omega^2 q^4/32 \pi^2 \varepsilon^2)^{1/3}$, respectively.

To find the ground and metastable states, we used classical molecular dynamics (MD) together with a suitable
``simulated annealing'' method. Starting with a random initial configuration of $N$ particles, the system is
cooled continuously until all momenta are zero and the particles settle in minima of the potential energy surface.
Depending on the particle number, the cooling down process was repeated
between a several hundred and a several thousand times until every of the computed low energy states
was found more than a given number of times (typically $10 \dots 100$) assuring a high probability
(though no general proof)
that also the ground state has been found. Crucial for a high search efficiency is
the use of an optimized MD time step (it has to be chosen not too small to avoid trapping in
local potential minima). The results are shown in tables \ref{tab:groundstates}
and \ref{tab:groundstates2}.

\begin{figure}
\includegraphics[angle=-90,width=19.1cm]{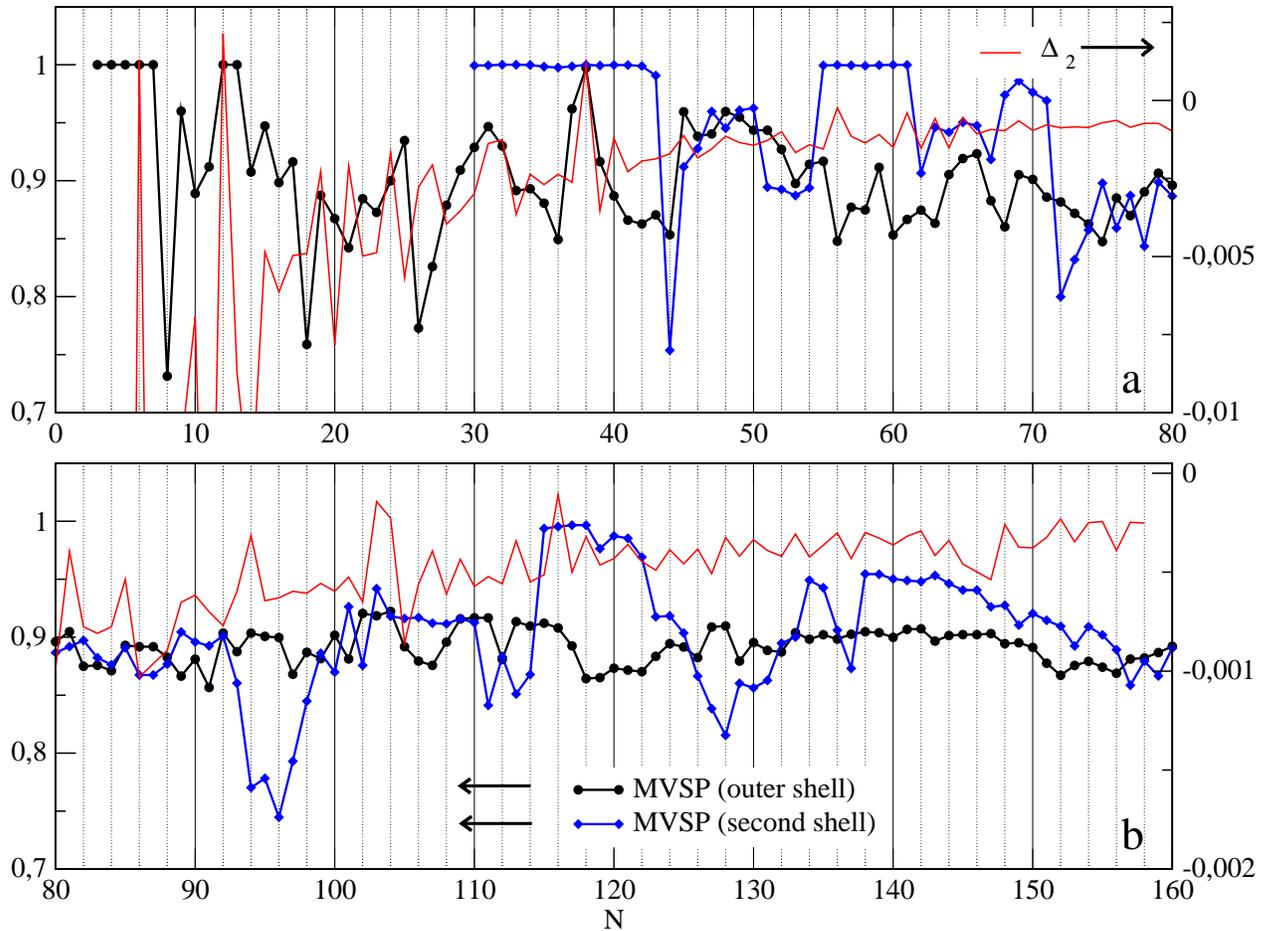}
\caption{\label{fig0} Binding energy $\Delta_2$ (right axis) and mean Voronoi symmetry parameter
(MVSP, left axis) for the two outermost cluster shells. {\bf a}: $N\le 80$,  {\bf b}: $80\le N\le 160$
(Color online).
}
\end{figure}

\begin{table}
\caption{\label{tab:groundstates}Shell
configurations, energy per particle for the lowest lying states (for the
excited states the \textit{energy difference} with respect to the ground
state is given), mean radius of outer shell
$r_{1}$, symmetry parameter $G_M$ and number of Voronoi M-polygons
$N(M)$ in brackets. For $N=4, N(3)=4$, and for $N=5, N(3)=2, N(4)=3$.}
\begin{ruledtabular}
\begin{tabular}[t]{|c|c|c|c|c|c|c|}
$N$  & Config.    & $E/N$     & $r_1$   & $G_4 [N(4)]$& $G_5 [N(5)]$& $G_6 [N(6)]$\\ [0.3ex]
\hline
2    & (2)        & 0.750000  & 0.5000  &   --        &   --       &   --            \\ [0.2ex]
\hline
3    & (3)        & 1.310371  & 0.6609  &   --        &   --       &   --            \\ [0.2ex]
\hline
4    & (4)        & 1.785826  & 0.7715  &   --        &   --       &   --            \\ [0.2ex]
\hline
5    & (5)        & 2.245187  & 0.8651  &   1.000 [3] &   --       &   --            \\ [0.2ex]
\hline
6    & (6)        & 2.654039  & 0.9406  &   1.000 [6] &   --       &   --            \\ [0.2ex]
\hline
7    & (7)        & 3.064186  & 1.0106  &   1.000 [5] &  1.000 [2] &   --            \\ [0.2ex]
\hline
8    & (8)        & 3.443409  & 1.0714  &   0.641 [4] &  0.821 [4] &   --            \\ [0.2ex]
\hline
9    & (9)        & 3.809782  & 1.1269  &   0.965 [3] &  0.957 [6] &   --            \\ [0.2ex]
\hline
10   & (10)       & 4.164990  & 1.1783  &   1.000 [2] &  0.861 [8] &   --            \\ [0.2ex]
     & (9,1)      & \textit{0.021989}  & 1.2453  &   0.965 [3] &  0.957 [6] &   --            \\ [0.2ex]
\hline
11   & (11)       & 4.513275  & 1.2265  &   0.940 [2] &  0.894 [8] &   1.000 [1]     \\ [0.2ex]
     & (10,1)     & \textit{0.009876}  & 1.2878  &   1.000 [2] &  0.861 [8] &   --             \\ [0.2ex]
\hline
12   & (12)       & 4.838966  & 1.2700  &   --        &  1.000 [12]&   --             \\ [0.2ex]
     & (11,1)     & \textit{0.015345}  & 1.3286  &   0.938 [2] &  0.895 [8] &   1.000 [1]     \\ [0.2ex]
\hline
13   & (12,1)     & 5.166798  & 1.3659  &   --        &  1.000 [12]&   --             \\ [0.2ex]
     & (13)       & \textit{0.005061}  & 1.3130  &   1.000 [1] &  0.894 [10] &   0.932 [2]    \\ [0.2ex]
\hline
14   & (13,1)     & 5.485915  & 1.4033  &   1.000 [1] &  0.893 [10] &   0.933 [2]    \\ [0.2ex]
     & (14)       & \textit{0.003501}  & 1.3527  &   --        &  0.938 [12] &   1.000 [2]    \\ [0.2ex]
\hline
15   & (14,1)     & 5.792094  & 1.4383  &   --        &  0.938 [12] &   1.000 [2]    \\ [0.2ex]
     & (15)       & \textit{0.009031}  & 1.3906  &   --        &  0.885 [12] &   0.963 [3]    \\ [0.2ex]
\hline
16   & (15,1)     & 6.093421  & 1.4719  &   --        &  0.882 [12] &   0.962 [3]    \\ [0.2ex]
     & (16)       & \textit{0.012200}  & 1.4266  &   --        &  0.897 [12] &   0.993 [4]    \\ [0.2ex]
     & (16)       & \textit{0.012635}  & 1.4267  &   --        &  0.747 [12] &   0.884 [4]    \\ [0.2ex]
\hline
17   & (16,1)     & 6.388610  & 1.5042  &   --        &  0.891 [12] &   0.993 [4]    \\ [0.2ex]
     & (16,1)     & \textit{0.000365}  & 1.5042  &   --        &  0.746 [12] &   0.884 [4]    \\ [0.2ex]
     & (17)       & \textit{0.015766}  & 1.4611  &   --        &  0.738 [12] &   0.810 [5]    \\ [0.2ex]
\hline
18   & (17,1)     & 6.678830  & 1.5353  &   --        &  0.738 [12] &   0.810 [5]    \\ [0.2ex]
     & (18)       & \textit{0.018611}  & 1.4941  &   1.000 [2] &  0.829 [8]  &   0.920 [8]    \\ [0.2ex]
\hline
19   & (18,1)     & 6.964146  & 1.5654  &   1.000 [2] &  0.827 [8]  &   0.920 [8]    \\ [0.2ex]
\hline
20   & (19,1)     & 7.247181  & 1.5946  &   --        &  0.838 [12] &   0.918 [7]    \\ [0.2ex]
     & (18,2)     & \textit{0.004264}  & 1.6285  &   0.991 [2] &  0.824 [8]  &   0.913 [8]    \\ [0.2ex]
\hline
21   & (20,1)     & 7.522378  & 1.6226  &   --        &  0.792 [12] &   0.917 [8]    \\ [0.2ex]
     & (19,2)     & \textit{0.004668}  & 1.6557  &   --        &  0.847 [12] &   0.927 [7]    \\ [0.2ex]
\hline
22   & (21,1)     & 7.795469  & 1.6499  &   1.000 [1] &  0.877 [10] &   0.880 [10]   \\ [0.2ex]
     & (21,1)     & $2.5\cdot 10^{-7}$  & 1.6499  &   1.000 [1] &  0.859 [10] &   0.866 [10]   \\ [0.2ex]
     & (20,2)     & \textit{0.000976}  & 1.6821  &   --        &  0.801 [12] &    0.935 [8]   \\ [0.2ex]
     & (20,2)     & \textit{0.001053}  & 1.6820  &   --        &  0.763 [12] &    0.909 [8]   \\ [0.2ex]
\end{tabular}
\end{ruledtabular}
\end{table}

Consider first the ground state shell configurations beyond the previously reported results
\cite{rafac91,tsuruta93}, see Tab. \ref{tab:groundstates2}. Closure of the second shell is observed twice:
for $N=57$ \cite{tsuruta93} and $N=60$. Further, we have found the closure of the third shell to occur at $N=154$, all larger clusters have at least four shells (in the ground state).
The ``nobel gas like'' closed shell clusters are particularly stable, but a few others also have a comparatively
high binding energy (addition energy change) $\Delta_2(N)=E(N+1)+E(N-1)-2E(N)$. Tsuruta et al. \cite{tsuruta93}
found the stable clusters $N=4, 6, 10, 12, 19, 32, 38, 56$. For larger clusters the binding energy decreases,
and the relative differences  rapidly decrease. We found the next particularly stable ones to be $N=81, 94, 103, 116$. The results are shown in Fig.~\ref{fig0}.
The relative stability of these clusters is linked to a particularly symmetric particle arrangement within the shells which will be analyzed below.

The existence of the shell structure is a marked difference from macroscopic Coulomb systems ($N \rightarrow \infty$) and is, of course, caused by the spherical confinement potential. With increasing $N$ the structure of a macroscopic system emerges gradually, see also Ref.~\cite{schiffer02}.
This can be seen from the relative widths ${\bar \sigma_m}\equiv \sigma_m/r_m$ of the $m-$th shell
($\sigma_m$ denotes the variance of the shell radius $r_m$).
For example, for $N=149$, (starting from the outermost shell)
${\bar \sigma_1}=0.0089$, ${\bar \sigma_2}=0.035$ and ${\bar \sigma_3}=0.032$,
whereas for $N=160$ we obtain ${\bar \sigma_1}=0.0091$, ${\bar \sigma_2}=0.033$ and
${\bar \sigma_3}=0.0038$. In both cases the outermost shell is significantly narrower than the second one and this trend becomes more pronounced as $N$ increases. This is easy to understand
because the effect of the confinement is strongest at the cluster boundary, i.e. in the outer
shell, whereas the inner shells are partially shielded from the trap potential by the surrounding particle shells. In contrast, the behavior of the inner shells is not that systematic: in one case ($N=149$) the third shell is of similar (relative) width as the second, in the other case ($N=160)$ the inner shell is much narrower. The reason are symmetry effects which particularly strongly influence the
width of the innermost shell (the cluster $N=160$ has a closed inner shell with 12 particles which is very narrow). 

In Tab. \ref{tab:groundstates} we also provide the first excited states which correspond to metastable shell configurations which are different from the ground state. While the overall trend is a rapid decrease of the
excitation energy (energy gap to the ground state) with increasing $N$, some additional systematics is observed.
Clusters which open a new shell typically possess a close metastable state. For example, for $N=13$ the
relative stability of the configurations $\{N,0\}$ and $\{N-1,1\}$ changes, the latter becomes the ground state and the former the first excited state, cf. Tab.\ref{tab:groundstates}. A similar trend is observed not only when a new shell is opened
but also when an additional particle moves onto the inner shell between the states $\{N_1-1,N_2\}$ and $\{N_1,N_2-1\}$. Away from these transition points the energy differenc increases and eventually another configuration becomes the first excited state.

\begin{figure}
\includegraphics[angle=-90,width=18cm]{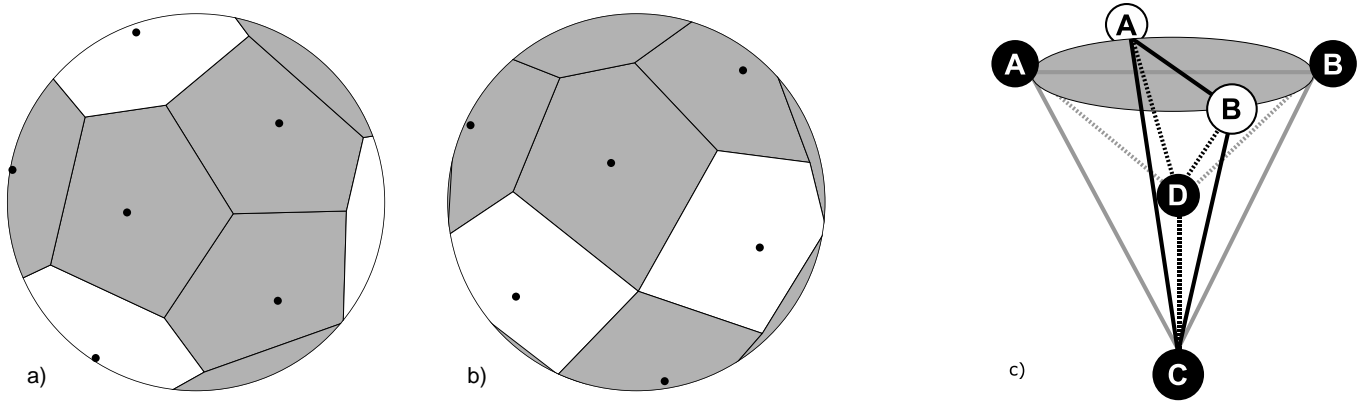}
\caption{\label{fig1} Voronoi construction for the cluster $N=17$ -- the two energetically lowest states with shell configuration $N=\{1,16\}$. White (grey) areas are hexagons (pentagons) -- indicating
the number of nearest neighbors of the corresponding particle (black dot). {\bf a}: ground state, {\bf b}: first excited (``fine structure'') state, {\bf c}: arrangement of
the four particles surrounded by hexagons -- the two states differ by rotation of the edge $AB$,
black [white] circles correspond to case a) [b)].
}
\end{figure}

An interesting observation
is that frequently simulations yielding the same shell configuration resulted in different total energies,
see e.g. $N=16, 17, 22$ in Tab. \ref{tab:groundstates}. The differences are
much larger than the simulation error, moreover, the energies are reproducible. The obvious explanation is that
the state of a cluster is not completely determined by its shell configuration (as
it is the case in 2D). In addition, there exist further excited states, i.e. a ``fine structure'', which are due to
a {\em different particle arrangement and symmetry within one shell}.
To understand the differences in the structure of these states with same shell configuration we analyzed the
intrashell symmetry by performing a Voronoi analysis, i.e. by constructing polygons around a given particle formed
by the lines equally bisecting nearest-neighbor pairs on the shell, cf. the example of $N=17$ shown in
Fig.~\ref{fig1}. Interestingly, both states do not differ with respect to the number of polygons of each kind
on the outer shell:
there are $N(5)=12$ pentagons and $N(6)=4$ hexagons. However, what is different is the
{\em arrangement of the polygons}: in one
case, the four hexagons form a perfect tetrahedron ABCD and are separated from each other by pentagons,
cf. Fig.~\ref{fig1}.a,
in the other two pairs of hexagons touch, see Fig.~\ref{fig1}.b, and the tetrahedron is distorted,
as shown in Fig.~\ref{fig1}.c. Two edges remain practically constant
($\overline{AB} \approx \overline{CD}\approx 1.63$), but the edge AB rotates with respect to the first case
by an angle of 34 degrees resulting in a reduction of edges $\overline{BC}$ and $\overline{AD}$ to about $1.24$
while $\overline{AC}$ and $\overline{BD}$ increase to $1.94$. From this we conclude that of two states the one
with the more symmetric arrangement of the Voronoi polygons, i.e. (Fig.~\ref{fig1}.a), has the lower energy.
To quantify this topological criterion, we introduce the {\em Voronoi symmetry parameter}
\begin{equation}\label{eq:order_parameter}
G_M = \frac{1}{N_M}\sum_{j=1}^{N_M}
  \frac{1}{M}\left| \sum_{k=1}^{M}e^{i M \theta_{jk}}
  \right|,
\end{equation}
where $N_M$ denotes the number of all particles $j$ in the shell, each of which is surrounded by
a Voronoi polygon of order $M$, ($M$ nearest neighbors) and $\theta_{jk}$
is the angle between the $j$th particle and its $k$th nearest neighbor. A value $G_5=1$ ($G_6=1$)
means that all pentagons (hexagons) are perfect, the magnitude of the reduction of $G_M$ below $1$
measures their distortion. Inspection of the values of $G_M$ for the two $\{16,1\}$ configurations
for $N=17$  (Tab. \ref{tab:groundstates}) reveals that the state with lower energy has higher values
for both $G_5$ and $G_6$ than the second, confirming our observation above. This result is verified for
all other $N$ (of course it applies only to states with the same shell configuration).

\begin{table}
\caption{\label{tab:groundstates2}Shell
configurations, energy per particle for the lowest lying states, and
 mean shell radii $r_{1,2,3}$ \cite{complete}}
\begin{ruledtabular}
\begin{tabular}[t]{|c|c|c|c|c|c|}
$N$ & Config.       & $E/N$     & $r_{1}$ & $r_{2}$ & $r_{3}$\\ [0.3ex] \hline
28  & (25,3)        & 9.348368  & 1.8525  & 0.6889 & -- \\ [0.2ex] 
29  & (25,4)        & 9.595435  & 1.8992  & 0.7987 & -- \\ [0.2ex] 
30  & (26,4)        & 9.838965  & 1.9198  & 0.7961 & -- \\ [0.2ex] 
31  & (27,4)        & 10.079511 & 1.9399  & 0.7926 & -- \\ [0.2ex] 
44  & (36,8)        & 13.020078 & 2.2454  & 1.0845 & -- \\ [0.2ex] \hline
54  & (44,10)       & 15.085703 & 2.4186  & 1.1872 & -- \\ [0.2ex] 
55  & (43,12)       & 15.284703 & 2.4618  & 1.2772 & -- \\ [0.2ex] 
56  & (44,12)       & 15.482144 & 2.4743  & 1.2770 & -- \\ [0.2ex] 
57  & (45,12)       & 15.679350 & 2.4869  & 1.2763 & -- \\ [0.2ex] 
58  & (45,12,1)     & 15.875406 & 2.5126  & 1.3765 & -- \\ [0.2ex] 
59  & (46,12,1)     & 16.070103 & 2.5247  & 1.3764 & -- \\ [0.2ex] 
60  & (48,12)       & 16.263707 & 2.5236  & 1.2754 & -- \\ [0.2ex] 
64  & (49,14,1)     & 17.027289 & 2.6101  & 1.4478 & -- \\ [0.2ex] 
65  & (50,14,1)     & 17.215361 & 2.6212  & 1.4477 & -- \\ [0.2ex] \hline
80  & (60,19,1)     & 19.936690 & 2.8369  & 1.6002 & -- \\ [0.2ex] 
84  & (61,21,2)     & 20.632759 & 2.9064  & 1.7140 & 0.5426 \\ [0.2ex] 
94  & (67,24,3)     & 22.325841 & 3.0347  & 1.8356 & 0.7001 \\ [0.2ex] 
95  & (67,24,4)     & 22.491878 & 3.0522  & 1.8848 & 0.8089 \\ [0.2ex] 
96  & (68,24,4)     & 22.657271 & 3.0606  & 1.8846 & 0.8083 \\ [0.2ex] 
97  & (69,24,4)     & 22.822032 & 3.0687  & 1.8849 & 0.8095 \\ [0.2ex] 
98  & (69,25,4)     & 22.986199 & 3.0864  & 1.9055 & 0.8081 \\ [0.2ex] 
99  & (70,25,4)     & 23.149758 & 3.0945  & 1.9056 & 0.8071 \\ [0.2ex] 
100 & (70,26,4)     & 23.312759 & 3.1117  & 1.9259 & 0.8055 \\ [0.2ex] 
101 & (70,27,4)     & 23.475164 & 3.1291  & 1.9450 & 0.8028 \\ [0.2ex] \hline
103 & (72,27,4)     & 23.798274 & 3.1451  & 1.9443 & 0.8017 \\ [0.2ex] 
105 & (73,28,4)     & 24.120223 & 3.1696  & 1.9641 & 0.8020 \\ [0.2ex] 
107 & (75,28,4)     & 24.439666 & 3.1850  & 1.9640 & 0.8011 \\ [0.2ex] 
109 & (77,28,4)     & 24.757151 & 3.2005  & 1.9638 & 0.8006 \\ [0.2ex] 
111 & (77,29,5)     & 25.072584 & 3.2322  & 2.0249 & 0.8968 \\ [0.2ex] 
113 & (77,30,6)     & 25.385842 & 3.2637  & 2.0831 & 0.9640 \\ [0.2ex] 
115 & (77,32,6)     & 25.697308 & 3.2949  & 2.1162 & 0.9630 \\ [0.2ex] 
117 & (79,32,6)     & 26.007089 & 3.3094  & 2.1158 & 0.9622 \\ [0.2ex] 
119 & (81,32,6)     & 26.315442 & 3.3237  & 2.1156 & 0.9624 \\ [0.2ex] 
121 & (83,32,6)     & 26.622118 & 3.3379  & 2.1154 & 0.9614 \\ [0.2ex] 
123 & (83,34,6)     & 26.927195 & 3.3672  & 2.1493 & 0.9625 \\ [0.2ex] \hline
125 & (84,34,7)     & 27.230458 & 3.3884  & 2.1850 & 1.0340 \\ [0.2ex] 
128 & (85,35,8)     & 27.682123 & 3.4235  & 2.2358 & 1.0922 \\ [0.2ex] 
130 & (86,36,8)     & 27.981234 & 3.4445  & 2.2501 & 1.0917 \\ [0.2ex] 
133 & (88,37,8)     & 28.427062 & 3.4718  & 2.2642 & 1.0912 \\ [0.2ex] 
135 & (88,38,9)     & 28.722421 & 3.4992  & 2.3110 & 1.1436 \\ [0.2ex] 
137 & (90,38,9)     & 29.016328 & 3.5119  & 2.3110 & 1.1440 \\ [0.2ex] 
139 & (91,39,9)     & 29.308774 & 3.5316  & 2.3251 & 1.1430 \\ [0.2ex] 
141 & (92,40,9)     & 29.599900 & 3.5514  & 2.3387 & 1.1417 \\ [0.2ex] 
143 & (93,40,10)    & 29.889733 & 3.5707  & 2.3689 & 1.1932 \\ [0.2ex] 
145 & (94,41,10)    & 30.178106 & 3.5898  & 2.3825 & 1.1920 \\ [0.2ex] \hline
147 & (95,42,10)    & 30.465219 & 3.6087  & 2.3957 & 1.1923 \\ [0.2ex] 
149 & (96,43,10)    & 30.750998 & 3.6273  & 2.4090 & 1.1926 \\ [0.2ex] 
151 & (96,43,12)    & 31.035390 & 3.6524  & 2.4659 & 1.2814 \\ [0.2ex] 
153 & (97,44,12)    & 31.318528 & 3.6708  & 2.4781 & 1.2811 \\ [0.2ex] 
154 & (98,44,12)    & 31.459632 & 3.6768  & 2.4777 & 1.2810 \\ [0.2ex] 
155 & (98,44,12,1)  & 31.600488 & 3.6887  & 2.5042 & 1.3846 \\ [0.2ex] 
156 & (98,45,12,1)  & 31.741100 & 3.7006  & 2.5169 & 1.3839 \\ [0.2ex] 
158 & (100,45,12,1) & 32.021294 & 3.7122  & 2.5166 & 1.3834 \\ [0.2ex] 
160 & (102,45,12,1) & 32.300405 & 3.7238  & 2.5161 & 1.3833 \\ [0.2ex]
\end{tabular}
\end{ruledtabular}
\end{table}

Having obtained with $G_M$ a suitable symmetry measure which is sensitive to the relative stability of
ground and metastable states, we now return to the issue of the overall
cluster stability. To this end we compute the {\em mean Voronoi symmetry parameter} (MVSP)
by averaging over all $G_M$ of a given shell weighted with the respective particle numbers $N(M)$. The results for the two outer shells for $N\le 160$
are included in Fig.~\ref{fig0}. We clearly see that {\em magic clusters} have not only a high binding energy
but also a prominent symmetry \cite{tsuruta93}, see in particular $N=12$, $N=38$, $N=103$ and $N=116$.


In summary, in this paper we have presented extensive simulation results for spherical Coulomb clusters with
$N\le 160$. The observed lowest energy states for $N\ge 60$ are, in most cases, lower than those previously reported
and should be valuable reference for experiments with classical Coulomb balls in dusty plasmas or ultracold ions.
Moreover, the ground state results (shell configurations) are expected to be important
also for quantum Coulomb clusters (e.g. in quantum dots) in the strong coupling
limit, as for 2D systems it was found that, in most cases, they have the same shell
configuration as their classical counterpart \cite{afilinov-etal.01prl,ludwig03}.
Further, we have presented an analysis of the lowest excited states of small clusters. Besides metastable states
with a shell structure different from the ground state we identified ``fine structure'' states
which are characterized by different particle arrangement within the shells. These states have a lower symmetry
which is linked to higher values of the total energy.
Despite the decreasing values of the excitation energy with increasing $N$, knowledge of the lowest metastable
states is important for understanding the dynamic properties of mesoscopic clusters.
We expect that the collective excitations of the clusters, i.e. the normal modes which are excited in the
system if kinetic energy is supplied will be strongly influenced by the metastable states.
Further, these states sthould be of importance for the melting behavior of mesoscopic Coulomb balls.

\begin{acknowledgments}
The authors thank A. Piel and D. Block for stimulating discussions and V. Golubnychiy for
assistence with the figures. This work is supported by the Deutsche Forschungsgemeinschaft
under grant BO-1366/5 and CPU time at the Rostock Linux Cluster ``Fermion''.
\end{acknowledgments}

\end{document}